\documentstyle[pre,aps,multicol]{revtex}
\begin{document}

\title{Higher order Shapiro steps in ac-driven Josephson junctions}
\author{Kim {\O}.\ Rasmussen$^1$,
Vadim Zharnitsky$^1$\footnote{Present address: Division of Applied Mathematics, 
Brown University, Providence, RI 02912.},
Igor Mitkov$^{1,2}$, and Niels Gr{\o}nbech-Jensen$^1$}
\address{$^1$Theoretical Division and Center for Nonlinear Studies,\\
Los Alamos National Laboratory, Los Alamos, New Mexico 87545.\\
$^2$Applied Theoretical and Computational Physics Division
and Center for Nonlinear Studies,\\
Los Alamos National Laboratory, Los Alamos, New Mexico 87545.}
\date{\today}
\maketitle
\begin{abstract}
We demonstrate that the well known phase-locking mechanism leading to
Shapiro steps in ac-driven Josephson junctions is always accompanied by a
higher order phase-locking mechanism similar to that of the
parametrically driven pendulum. This effect, resulting in a $\pi$-periodic
effective potential for the phase, manifests itself clearly in the parameter
regions where the usual Shapiro steps are expected to vanish.
\end{abstract}
\begin{multicols}{2}
The harmonically driven pendulum has been studied extensively over the past
decades in many different contexts. The driven pendulum has been one of the key
systems in nonlinear dynamics due to its simplicity and richness in
nonlinear phenomena such as phase-locking \cite{VanDuzer,Kautz1,Miles,Jensen1}
and chaos \cite{Huberman,Kautz2,Jensen2}.
Another reason for studying the driven pendulum is that the
pendulum equation is the most widely used model for superconducting Josephson
junctions \cite{VanDuzer,Barone}.
Particularly phase-locking of pendulum motion to an
ac-perturbation has been extensively studied in the literature due to the
general interests in synchronization of oscillators and specifically due to
the related Josephson junction, where phase-locking of the voltage response
to an ac-current can occur in certain regions of the parameter space
\cite{Shapiro}.
There are two distinctively different ways of driving a pendulum;
direct and parametric drive. The direct (torsional) ac-drive, which we will
study in this paper, is the most relevant for technologically interesting
systems
such as the Josephson junction, and it has been found to produce efficient
harmonic \cite{VanDuzer,Shapiro} and subharmonic \cite{Filatrella}
phase-locking of great interest for, e.g., the voltage standard \cite{Kautz}.
The parametric drive (e.g., pivot oscillations) has mainly been studied
from the point of view of nonlinear dynamics. The usual theoretical technique
for predicting and analyzing the ac-driven pendulum is to assume that the
driving frequency is much larger than the natural oscillations of the
unperturbed pendulum. Within this framework, one can develop time separation
analyses separating a fast linear response to the ac-drive from the overall slow
nonlinear behavior of the system. The main difference between the
effects of the two driving mechanisms is that the parametric drive usually
leads to a $\pi$-periodic effective potential for the slow nonlinear behavior
\cite{Landau,Blackburn1,Vadim0}, while the direct drive leads to the usual
$2\pi$-periodicity \cite{VanDuzer,Barone}.
We will demonstrate in this paper that the direct ac-drive may lead to a
$\pi$-periodic effective potential in certain regions of the parameter space
and that this result has consequences for the well known Shapiro steps in the
current-voltage (IV) characteristics of ac-driven Josephson junctions.

We study the pendulum equation in the form,
\begin{eqnarray}
\ddot{\phi}+\alpha\dot{\phi}+\sin\phi & = & \eta + \varepsilon\sin{\Omega t}
\; ,
\label{eq:Eq_1}
\end{eqnarray}
where $\phi$ is the pendulum angle relative to vertical (down) and $\alpha$ is
a normalized friction coefficient. Time is normalized to $\tau=\sqrt{g/l}$,
where $g$ is the gravitational constant and $l$ is the length from the pivot
to the pendulum bob, $\eta$ is a dc-torque, normalized to $mgl$, $m$ being the
mass of the bob, and the normalized frequency, $\Omega$, and amplitude,
$\varepsilon$, define the direct ac-drive of the pendulum. In the context of
Josephson junctions, $\phi$ is the phase difference between the quantum
mechanical wave functions of the superconductors defining the junction,
$\alpha$ is the normalized dissipation coefficient due to transport of
quasiparticles, $\eta$ and $\varepsilon$ are currents normalized to the
critical current of the junction, and time is normalized to the inverse
Josephson plasma frequency, $\tau=\sqrt{\hbar C/2e I_c}$, where $\hbar$ is
Planck's constant, $C$ is the device capacitance, and $I_c$ is the critical
current. Voltage across the Josephson device is given by
$V=\dot\phi\hbar/2e$, uniquely relating Josephson voltage to
pendulum speed \cite{Barone}.

Let us write the phase in the following form,
\begin{eqnarray}
\phi & = & \theta + \xi t + \Xi(t) \; ,
\label{eq:Eq_2}
\end{eqnarray}
where $\Xi(t)$ is a function that oscillates with frequency $\Omega$,
$\xi$ is a constant to be determined, and $\theta$ is a phase,
$\langle\dot{\theta}\rangle=0$. Inserting
Eq.\ (\ref{eq:Eq_2}) into Eq.\ (\ref{eq:Eq_1}) we obtain,
\begin{eqnarray}
\ddot{\theta} + \ddot{\Xi} + \alpha\dot{\theta} + \alpha\xi + \alpha\dot{\Xi}
+ \sin\left(\theta+\xi t + \Xi\right) & = & \eta + \varepsilon\sin{\Omega t} \; .
\nonumber \\
\label{eq:Eq_3}
\end{eqnarray}
We will choose $\Xi$ so that, $\ddot{\Xi}+\alpha\dot{\Xi}=\varepsilon\sin{\Omega t}$,
and we then obtain,
\begin{eqnarray}
\Xi(t) & = & - \frac{\varepsilon}{\Omega\sqrt{\Omega^2+\alpha^2}}
\sin\left(\Omega t + \gamma\right) \; ,
\label{eq:Eq_4}
\end{eqnarray}
where $\gamma=\tan^{-1}\left(\alpha/\Omega\right)$ is a constant phase. Equation
(\ref{eq:Eq_3}) then takes the form,
\begin{eqnarray}
\ddot{\theta} + \alpha\dot{\theta} + \sin\left(\theta+\xi t + \Xi\right)
& = & \eta - \alpha\xi \; .
\label{eq:Eq_5}
\end{eqnarray}
Rewriting Eq.\ (\ref{eq:Eq_1}) in the form of Eq.\ (\ref{eq:Eq_5}) demonstrates
how the direct ac drive can lead to parametric effects similar to those
of the parametrically driven pendulum equation.

We can now apply three consecutive transformations that will lead to
an equation in variables where time-dependent terms are of order $\Omega^{-3}$. 
We describe the general procedure as a generalization of the analysis presented
in Ref.\ \cite{Vadim}
(which goes back to Poincare, see, {\em e.g.}, Ref.\ \cite{arnold}) in the
Appendix.

To average the equation we rewrite it as a system of two ordinary differential
equations (ODEs) \cite{Note}
\begin{eqnarray}
&&\dot \theta = p \label{vadim1}  \\ 
&&\dot p = \eta - \alpha \xi - \alpha p - A \sin \theta - B \cos \theta \; ,
\nonumber
\end{eqnarray}
where $A$ and $B$ are given by,
\begin{eqnarray*}
A & = & \cos\left(\xi t + \Xi\right) \\
B & = & \sin\left(\xi t + \Xi\right) \; .
\end{eqnarray*}

Carrying out the procedure described in the Appendix three times 
and neglecting terms ${\cal{O}}(\Omega^{-3})$ or higher we obtain  
\begin{eqnarray}
&& \dot{\Theta} = P \label{eq:result_eq} \\
&&\dot P = \eta-\alpha\xi -\alpha P-
G_1\sin\left(\Theta + \delta_{G_1}\right) -
G_2\sin\left(2\Theta+\delta_{G_2}\right) ,
\nonumber
\end{eqnarray}
where
\begin{eqnarray}
G_1 & = & \left\{\begin{array}{lll}
-J_k(\Gamma) & , & \; \; {\rm if} \; \; \xi/\Omega=k \; \; {\rm integer} \\
0            & , & \; \; {\rm otherwise}
\end{array}\right.
\label{eq:Eq_C} \\
{G_2} & = & \left\{\begin{array}{lll}
-\sum_{n\neq k}\frac{J_n(\Gamma)J_{2k-n}(\Gamma)}{\Omega^2(n-k)^2} & , &
\; \; {\rm if} \; \; \xi/\Omega=k \; \; {\rm integer} \\
-\sum_n\frac{J_n(\Gamma)J_{1-n}(\Gamma)}{\Omega^2(n-\frac{1}{2})^2} & , &
\; \; {\rm if} \; \; \xi/\Omega=k=\frac{1}{2} \\
0 & , & \; \; {\rm otherwise}
\end{array}\right. \nonumber \\
\label{eq:Eq_D} \\
\Gamma & = & -\frac{\varepsilon}{\Omega\sqrt{\Omega^2+\alpha^2}} \; , \nonumber
\end{eqnarray}
and
\begin{eqnarray*}
\delta_{G_1} & = & \tan^{-1}\frac{\langle B \rangle}{ \langle A \rangle} \\
\delta_{G_2} & = & \tan^{-1} \frac
{2 \langle \{ B \}_{-1} \{ A \}_{-1} \rangle}
{ \langle \{ B \}_{-1}^2 \rangle -\langle \{ A \}_{-1}^2 \rangle } \; .
\end{eqnarray*}
Brackets, $\langle\cdots\rangle$, denote time average and $\left\{f\right\}$
is given by, $\{f \} = f-\langle f \rangle$, where $f$ is a periodic function.
The mean-zero antiderivative, $\left\{f\right\}_{-1}$, is defined as,
\begin{eqnarray*}
\left\{f\right\}_{-1} & = & \int\left\{f\right\}dt \; , \; \;
\langle\left\{f\right\}_{-1}\rangle \; = \; 0 \; .
\end{eqnarray*}

It is now clear that phase-locking of the pendulum motion can exist for
values of $\eta$ given by
\begin{eqnarray}
\left|\eta-\alpha\xi\right| < \left|{G_1}\sin\left(\Theta + \delta_{G_1}\right)
+{G_2}\sin\left(2\Theta + \delta_{G_2}\right)\right| \; .
\label{eq:Eq_lock}
\end{eqnarray}
For ${G_2}=0$ this leads directly to the well-known Bessel function expression
for the Shapiro steps in ac-driven Josephson junctions \cite{VanDuzer} and
this result is correct up to order ${\cal{O}}(\Omega^{-2})$. However, we find
that, for $G_1=0$, i.e., at every node of the Bessel function $J_k$, we have
phase-locking originating from the coefficient $G_2$ to the $\pi$-periodic
term in Eq.\ (\ref{eq:result_eq}). As a consequence, we may predict that the
Shapiro steps in the IV characteristics of ac-driven Josephson junctions do
{\it not} vanish for parameter values given by $J_k(\Gamma)=0$.

In order to demonstrate this we have performed numerical simulations of the
equation (\ref{eq:Eq_1}) and measured the ranges of phase-locking in $\eta$ as a
function of $\Omega$, $\varepsilon$, $k$, and $\alpha$. Figure 1a shows a
typical normalized IV ($\eta$,$\langle\dot\phi\rangle$) characteristic of
an ac-driven system and we use this to define the magnitude of the locking
range, $\Delta\eta_k$. The system parameters have here been chosen to
$\alpha=0.3$, $\Omega=3$, and $\Gamma=5$.
In Fig.\ 1b we show the magnitude of the locking range, $\Delta\eta_k$,
against $\varepsilon$ for $\Omega=3$, $\alpha=0.1$, and $k=1$. The solid line
represents the usual Bessel function prediction, $\Delta\eta_1=2|G_1|$,
and the markers represent the numerical simulations.
It is clear that as long as $G_1$ (the relevant Bessel function) is not close
to one of the nodes the comparison between the solid line and the markers
is good. However, from the inset we see that close to the node of $G_1$, we
observe a relatively large discrepancy, which is obviously due to the
correction from the $\pi$-periodic effective potential given by $G_2$.
The best parameter range to study the effect of the $\pi$-periodic effective
potential is therefore to choose parameters such that $G_1\approx 0$; i.e.,
for $J_k(\Gamma)=0$ when $k$ is an integer.

Figure 2 shows direct comparisons between numerical
simulations (markers: $\alpha=0.05$ closed, $\alpha=0.1$ open) and our
predictions (solid line), $\Delta\eta_k=2|G_2|$, for parameter values leading
to $G_1=0$. The comparisons are performed at the smallest nonzero value of
$\Gamma$ for which $J_k(\Gamma)=0$, and we show
comparisons for $k=0$ (figure 2a), $k=1$ (figure 2b), $k=2$ (figure 2c),
and $k=\frac{1}{2}$ (figure 2d) keeping $\Gamma$ constant for each figure.
Note that in the latter case, $k=\frac{1}{2}$, $G_1$ is always zero.
We have here, arbitrarily, chosen $J_{0}(\Gamma=2.4)\approx 0.0$.
It is obvious that our comparisons
demonstrate an excellent agreement between simulations and prediction of the
magnitude of the phase-locked region in $\eta$ for all the different parameter
values. The comparisons are performed in the frequency range between
$\Omega=1$ and $\Omega=40$ since driving frequencies smaller than $\Omega=2$
typically lead to low stability of the phase locked states (the analysis is
developed for high $\Omega$) and since phase-locking becomes impractical
to identify for frequencies larger than $\Omega=30$. Our data
shows a slight trend of overestimating the locking range for large frequencies.
We have identified this to be an artifact of numerically solving the pendulum
equation with discrete time. Particularly, we have chosen the time step for the
simulations to be a fraction of the period of the driving frequency, and thus,
when looking for extremely small phase-locked steps in the normalized IV
characteristics, we are finding an artificial phase-locking of the dynamics
to the temporal discretization.

We have demonstrated
that the $\pi$-periodic effective potentials can exist in the directly ac-driven
pendulum and we have further given a quantitatively correct estimate of the
significance of this effect.
The magnitude of the $\pi$-periodic effective potential suggests that the
phase-locking signature of the potential
should be directly observable not only in the driven pendulum
\cite{Blackburn,Blackburn2},
but also in ac-driven Josephson junctions. For relatively low
driving frequencies, $\Omega\approx 2$, we observe locking ranges in figure 2
of the order of $\Delta\eta\approx{0.1}$, indicating that a standard dc
current-voltage characteristic of a current ac-driven junction will indeed show
a significant locking range where the usual Bessel function amplitude would
suggest that locking is not possible. A particularly convenient choice of
parameters is to operate the system at the subharmonic resonance,
$k=\frac{1}{2}$, where $G_1$ is always zero.

This work was performed under the auspices of the U.S.\ Department of Energy.
V.Z.\ Acknowledges partial support from NSF grant DMS-9627721.
 

\begin{center}
{\bf\large Appendix}
\end{center}

Let us consider a
system of equations (\ref{vadim1}) written in vector form
\begin{eqnarray*}
\dot x = f(x,\tau), 
\end{eqnarray*}
where $\tau = \Omega t$, $x=(\theta,p)$, and $f=(f_1,f_2)$.
We will apply an averaging procedure as follows.

Let $x=x_1+ \Omega^{-1}h_1 (x_1,\tau)$ be the first transformation with, 
yet undefined $h_1$. This function is restricted to be periodic in $\tau$
so that the new variables are close to the old ones uniformly in time.
In the new variables the equation takes the form
\begin{eqnarray*}
(I+ \Omega^{-1} D_{x_1} h_1) \dot x_1 + h_{1 \tau} = 
f(x_1+ \Omega^{-1}h_1 (x_1,\tau), \tau),
\end{eqnarray*}
where $I$ is the unit matrix and $D_x$ denotes differentiation with respect
to the elements in $x$.
We obtain by Taylor expanding $f$,
\begin{eqnarray*}
(I+ \Omega^{-1} D_{x_1} h_1) \dot x_1 + h_{1 \tau} \\
= f(x_1, \tau) + \Omega^{-1}D_{x_1} f(x_1, \tau) h_1 + \cdots
\end{eqnarray*}
In order to eliminate the oscillatory part of $f(x_1, \Omega t)$ 
we choose $h_1=\{ f(x_1,\tau) \}_{-1}$ ($\{g \}=g-\langle g \rangle$ and
$\{g \}_{-1}$ is a mean-zero antiderivative of $\{g \}$) to obtain
\begin{eqnarray*}
\dot x_1 = \langle f \rangle(x_1) + \Omega^{-1} R_1 (x_1,\tau,\Omega^{-1}),
\end{eqnarray*} 
where $R_1$ is polynomial in $\Omega^{-1}$. 
The second transformation, given by $x_1=x_2+\Omega^{-2} h_2 (x_2,\tau)$
with $h_2=\{ R_1 (x_1,\tau,0)\}_{-1}$, moves time-dependence to second
order in $\Omega^{-1}$
\begin{eqnarray*}
\dot x_2 = \langle f \rangle (x_2) + \Omega^{-1} 
\langle R_1 \rangle (x_2,0) + \Omega^{-2} R_2 (x_2,\tau,\Omega^{-1}).
\end{eqnarray*} 
Continuing this procedure one can bring the system to the form
\begin{eqnarray*}
\dot x_n & = &
\langle f \rangle (x_n) + \Omega^{-1} \langle R_1 \rangle (x_n,0) +\\
&& \cdots + \Omega^{-n+1} \langle R_{n-1} \rangle (x_n,0)+ 
\Omega^{-n} R_n  (x_n,\tau,\Omega^{-1}).
\end{eqnarray*}
\end{multicols}
\begin{figure}
\vspace*{250pt}
\hspace{5.40 in}
\includegraphics{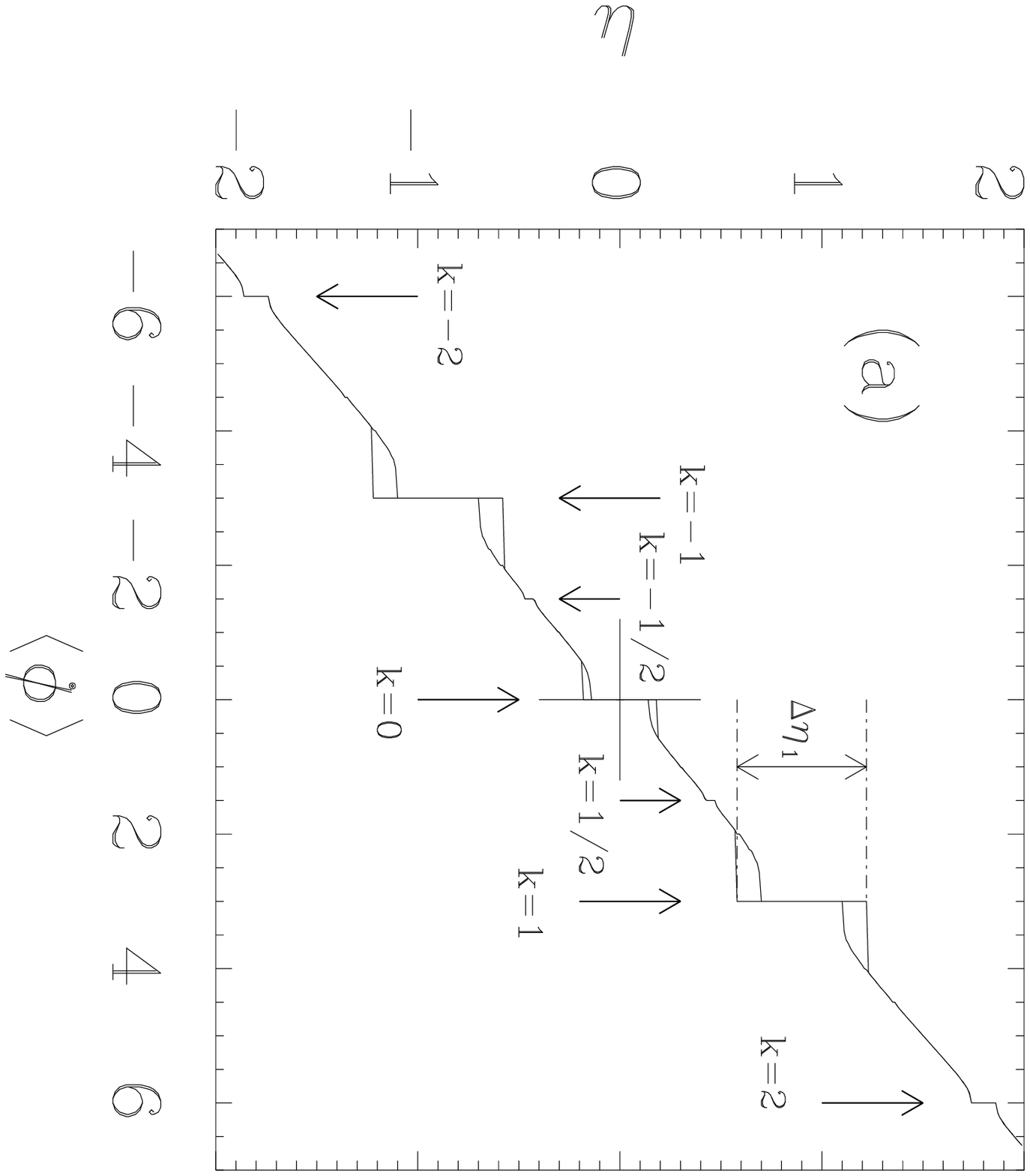}

\vspace{2.7 in}

\hspace{5.40 in}
\includegraphics{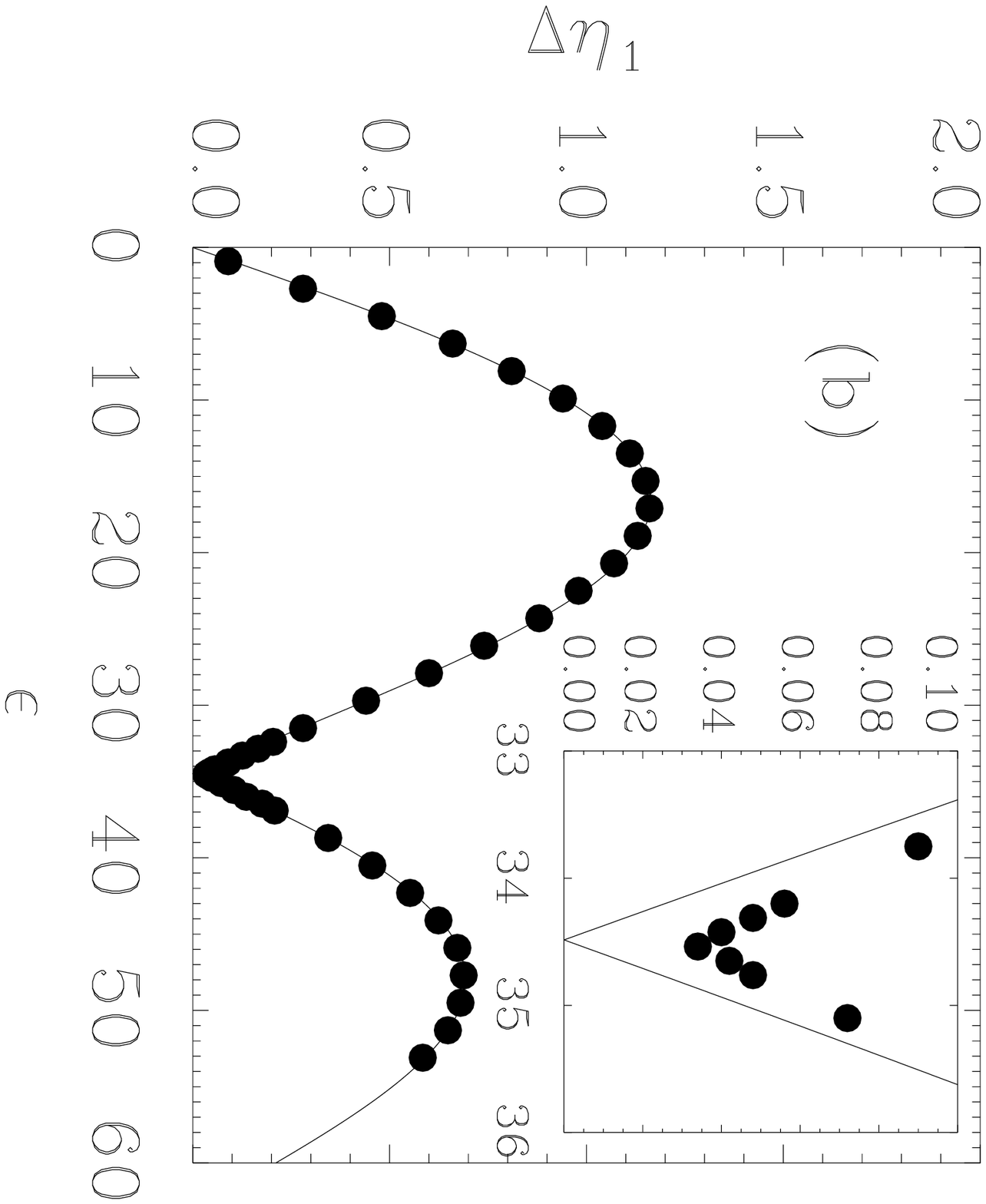}
\vspace*{-30pt}
\caption{(a) Simulated IV ($\eta$,$\langle\dot{\phi}\rangle$)
characteristics of the ac-driven system described by
Eq.\ (\ref{eq:Eq_1}). Parameters are: $\Omega=3$, $\alpha=0.3$, and $\Gamma=5$.
Relevant resonant $k$ steps are indicated with arrows, and the range of
phase-locking, $\Delta\eta_k$ is indicated for the $k=1$ step.
(b) Magnitude, $\Delta\eta_1$, of the phase-locked step as a function of
$\varepsilon$ for parameters:
$\Omega=3$ and $\alpha=0.1$. Solid line represents the usual prediction of
the Shapiro step, $\Delta\eta_1=2|G_1|$ for $k=1$ and markers are results of
numerical simulations. Inset shows details near the point $J_1(\Gamma)=0$.}
\end{figure}

\begin{figure}
\vspace*{300pt}
\hspace{6.00 in}
\includegraphics{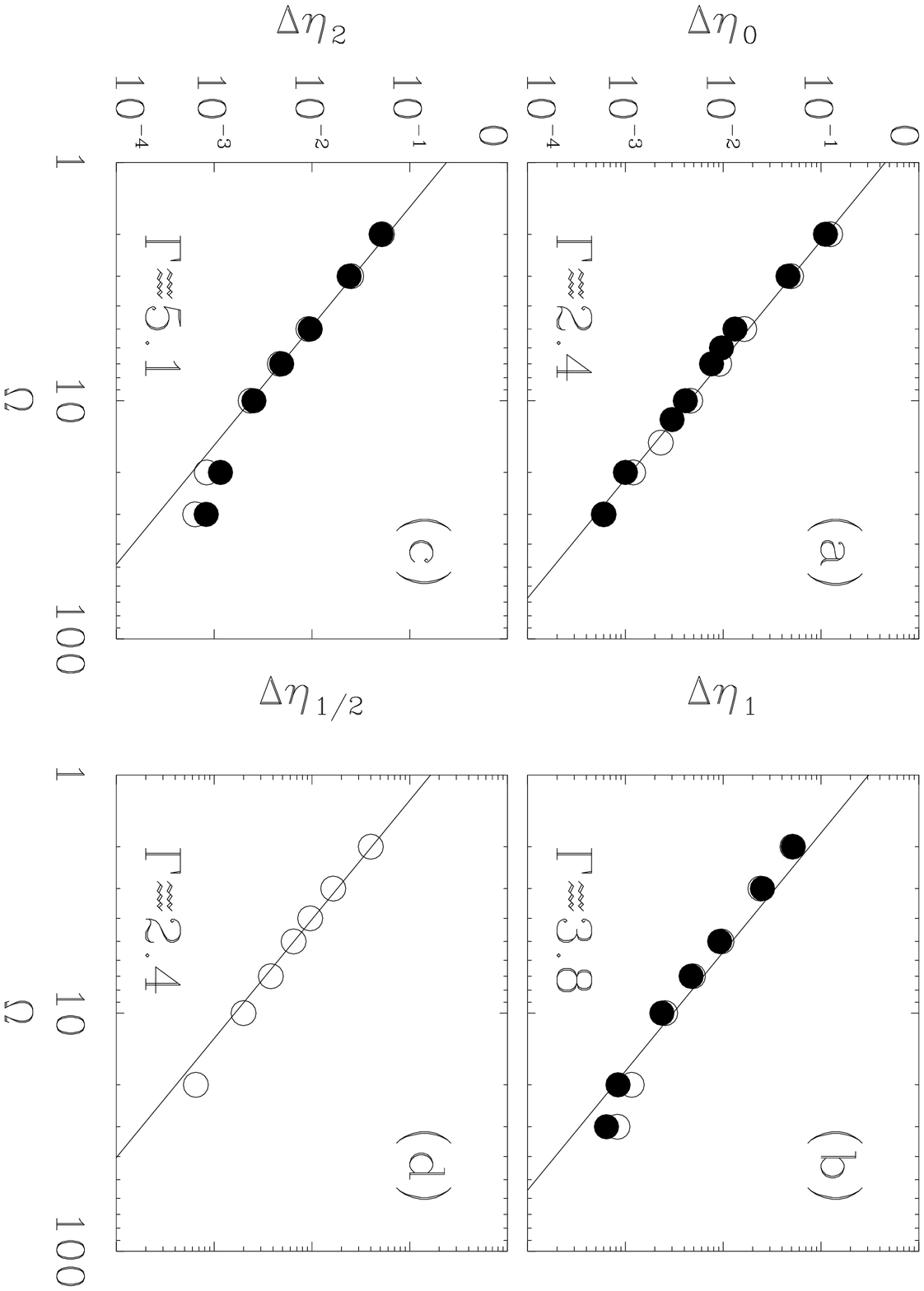}
\vspace*{-00pt}
\caption{Minimum of the locking range in $\eta$ as a function of the driving
frequency $\Omega$ near the first node of $G_1$
for nonzero $\Gamma$. Solid lines show the prediction, $\Delta\eta_k=2|G_2|$
and markers are results of numerical simulations of Eq.\ (\ref{eq:Eq_1}). Open
markers are for $\alpha=0.1$ and closed are for $\alpha=0.05$.
(a) $\Gamma\approx 2.4$ and $k=0$.
(b) $\Gamma\approx 3.8$ and $k=1$.
(c) $\Gamma\approx 5.1$ and $k=2$.
(d) $\Gamma\approx 2.4$ and $k=\frac{1}{2}$.}
\end{figure}

\begin{references}
\bibitem{Note} It is important to apply the normal form procedure, see the Appendix,
to the system of two first order ODEs for $(\theta,p)$ rather
than to the original second order ODE for $\theta$. The reason 
is that  the group of transformations which are used to move 
rapidly oscillating terms  to higher order must be 
sufficiently large. In the first case we obtain the averaged
equations by transforming $(\theta,p)$, see the Appendix.
In the second case it is impossible to arrive at the same result 
as the transformations of only one variable $\theta$ do not give enough 
freedom. This is another manifestation of the well known fact
that transformations in configuration
space form a subgroup of canonical transformations in phase space.
\end{references}
\end{document}